\documentclass{pasj01}
\usepackage{url}

\begin{document}
\SetRunningHead{Tsujimoto et al.}{Suzaku \& NuSTAR X-Ray Spectroscopy of $\gamma$\,Cas and HD\,110432}

\Received{\today}
\Accepted{\today}
\Published{}

\title{Suzaku and NuSTAR X-Ray Spectroscopy of $\gamma$\,Cas and HD\,110432}
\author{
Masahiro~\textsc{Tsujimoto},\altaffilmark{1}
Kumiko~\textsc{Morihana},\altaffilmark{2}
Takayuki~\textsc{Hayashi},\altaffilmark{2,3}
Takao~\textsc{Kitaguchi}\altaffilmark{4}
}
\altaffiltext{1}{Japan Aerospace Exploration Agency, Institute of Space and Astronautical Science, Chuo-ku, Sagamihara, Kanagawa 252-5210, Japan}
\email{tsujimot@astro.isas.jaxa.jp}
\altaffiltext{2}{Department of Physics, Nagoya University, Chikusa-ku, Nagoya, Aichi 464-8602, Japan}
\altaffiltext{3}{NASA's Goddard Space Flight Center, Greenbelt, MD 20771, USA}
\altaffiltext{4}{RIKEN Nishina Center, Wako, Saitama 351-0198, Japan}
\KeyWords{stars: individual ($\gamma$\,Cas, HD\,110432) --- stars: emission-line, Be --- stars: white dwarfs --- X-rays: stars}
\maketitle

\begin{abstract}
 $\gamma$\,Cas and its dozen analogs comprise a small but distinct class of X-ray
 sources. They are early B\textit{e}-type stars with an exceptionally hard thermal X-ray
 emission. The X-ray production mechanism has been under intense debate. Two competing
 ideas are (i) the magnetic activities in the B\textit{e} star and its disk and (ii) the
 mass accretion onto the unidentified white dwarf (WD). We adopt the latter as a working
 hypothesis and apply physical models developed to describe the X-ray spectra of
 classical WD binaries containing a late-type companion. Models of non-magnetic and
 magnetic accreting WDs were applied to $\gamma$ Cas and its brightest analog HD\,110432
 using the Suzaku and NuSTAR data. The spectra were fitted by the two models, including
 the Fe fluorescence and the Compton reflection in a consistent geometry. The derived
 physical parameters are in a reasonable range in comparison to their classical WD
 binary counterparts. Additional pieces of evidence in the X-ray spectra ---partial
 covering, Fe L lines, Fe \emissiontype{I} fluorescence--- were not conclusive enough to
 classify these two sources into a sub-class of accreting WD binaries. We discuss
 further observations, especially long-term temporal behaviors, which are important to
 elucidate the nature of these sources more if indeed they host accreting WDs.
\end{abstract}

\section{Introduction}\label{s1}
$\gamma$\, Cas is a B\textit{e} star with a spectral type B (B0.5 IV\textit{e}) showing,
at times, prominent emission lines including the Balmer series of H
\citep{jaschek81}. The emission is considered to originate from the disk, the shell, or
both around the rapidly spinning star with a significant mass loss. The source is
recognized as an archetype of B\textit{e} stars and is studied in great detail in the
near-infrared, optical, and ultra-violet bands.

In the X-rays, however, $\gamma$\,Cas is quite anomalous. Previous observations
\citep{mason76,frontera87,white82,murakami86,horaguchi94,haberl95,smith98a,kubo98,owens99,robinson00,robinson02,smith04,denhartog06,lopes10,shrader15,hamaguchi16}
revealed an X-ray luminosity of $\sim$10$^{32-33}$~erg~s$^{-1}$, the lack of drastic flux
variations such as X-ray outbursts, the presence of the Fe K complex resolved into three
emission lines from highly ionized Fe (Fe\emissiontype{XXV} He$\alpha$ and
Fe\emissiontype{XXVI} Ly$\alpha$ respectively at 6.7 and 7.0~keV) as well as
quasi-neutral Fe (Fe\emissiontype{I} K$\alpha$ at 6.4 keV). It has an exceptionally high
plasma temperature beyond $\sim$10~keV, which is unseen in other classes of early-type
stars. These distinct characteristics are shared by a dozen other sources, which are
called $\gamma$ Cas analogs\footnote{In this paper, we include $\gamma$ Cas itself in
$\gamma$ Cas analogs.} \citep{smith16,naze17}. The brightest source next to $\gamma$ Cas is
HD\,110432 \citep{smith06b,lopes07,torrejon12,smith12}. These two sources are used to
derive general characteristics of this enigmatic class of sources with their sufficient
brightness. We use this duo in the present study.

\medskip

Over 20 years, there has been an intensive but unsettled debate for the production
mechanism of the hot plasma responsible for the hard thermal X-rays
\citep{motch15,smith16}. Two competing ideas are (a) the magnetic activities of the
B\textit{e} star and its decretion disk \citep{robinson00} and (b) the accretion from
the B\textit{e} star to an unidentified white dwarf (WD; \cite{haberl95,kubo98}). The
purpose of this paper is to adopt the latter as a working hypothesis and to investigate
how much we can constrain the nature of WDs from X-ray spectroscopy if indeed $\gamma$
Cas analogs are B\textit{e}/WD binaries.

In fact, X-ray spectroscopy is one of the most powerful tools to reveal the properties
of WDs in classical WD binaries comprised of a WD and a late-type
star\footnote{In this paper, WD binaries are used for all systems including a WD and a
star. Classical WD binaries are used when we specifically refer to those with a
late-type companion in a semi-detached system.} \citep{mukai17}. A binary system is
called a cataclysmic binary if the companion is a dwarf and a symbiotic star if the
companion is a giant. Regardless of the spectral type and the luminosity class of the
companion, however, X-rays are produced close to the surface of WDs hence their spectral
shape is primarily governed by the properties of WDs and how the X-ray plasma is
fueled. Many physical models have been developed to describe X-ray spectra of different
sub-classes of classical WD binaries, which can constrain these properties.

The X-ray spectral modeling of $\gamma$ Cas analogs is far behind these developments. In
all previous work, the spectra were fitted only with phenomenological models comprised
of several components of iso-thermal collisionally-ionized plasma emission
\citep{white82,murakami86,horaguchi94,haberl95,kubo98,owens99,robinson00,smith04,shrader15,hamaguchi16,torrejon01,lopes07,torrejon12}. Under
the working hypothesis of this paper, we apply physical models developed for classical
WD binaries to the two brightest $\gamma$ Cas analogs. We then investigate if these two
sources can be interpreted reasonably by any or none of the sub-classes of WD binaries.

\medskip

The structure, the strategy, and the conclusion of this paper are as follows. In
\S~\ref{s2}, we describe the instruments that we use for constructing the X-ray spectra of
$\gamma$ Cas and HD\,110432. We choose the Suzaku and the Nuclear Spectroscopic
Telescope Array (NuSTAR) observatories because they cover a wide energy range including
the hard ($>$10~keV) band. The hottest component of the plasma, which is accessible in
the hard band, is produced by converting all the available energy into heat in accreting
WDs, hence it is most sensitive to the WD mass.

In \S~\ref{s3}, we present the result of the spectral analysis. We employ two physical
models routinely used to describe X-ray spectra of non-magnetic accreting WDs
(\S~\ref{s3-1}) and magnetic accreting WDs (\S~\ref{s3-2}). WD binaries fueled by the
nuclear fusion ---such as classical novae \citep{starrfield16} or super-soft sources (SSS;
\cite{kahabka97})--- do not share the spectral and temporal characteristics with any of
the $\gamma$ Cas analogs, so we do not consider these possibilities.

In \S~\ref{s4}, we discuss how much we learn by hypothesizing that $\gamma$ Cas analogs
are WD binaries. In \S~\ref{s4-1}, we compare the result of these sources with classical
WD binaries and show that they are reasonably interpreted within the range of accreting
WD binaries. In \S~\ref{s4-2}, we discuss whether the two $\gamma$ Cas analogs are
non-magnetic or magnetic. We examine additional pieces of evidence in the X-ray spectra,
but none of them are conclusive enough to judge whether the two sources are non-magnetic
or magnetic. In \S~\ref{s4-3}, we discuss that further observations, especially
long-term temporal behaviors, are important to elucidate the nature of these sources
more if indeed they host accreting WDs. The main results are summarized in
\S~\ref{s5}.

\section{Observations and Data Reduction}\label{s2}
\begin{table*}[hbtp]
 \caption{Observation log}\label{t01}
 \begin{center}
  \begin{tabular}{lccllccc}
   \hline
   \hline
   Object & RA & Dec & Observatory/Instrument & Sequence    & Observation date & $t_{\rm{exp}}$\footnotemark[$*$]& CR\footnotemark[$\dagger$]\\
          & \multicolumn{2}{c}{(J2000.0)} &                        & number      & (UT) & (ks) & (s$^{-1}$) \\
   \hline
   $\gamma$\,Cas & \timeform{00h56m38s} & \timeform{+60D44'08''} & Suzaku/XIS, PIN & 406040010   & 2011/07/13--14 & 55.4 & 11/0.59 \\
                 &                      &                        & NuSTAR/FPMA, B  & 30001147002 & 2014/07/24--25 & 31.0 & 6.5 \\
   HD\,110432    & \timeform{12h42m50s} & \timeform{-63D03'31''} & Suzaku/XIS, PIN & 403002010   & 2008/09/09--10 & 25.3 & 2.5/0.64 \\
   \hline
   \multicolumn{8}{@{}l@{}}{\hbox to 0pt{\parbox{180mm}{
   \footnotesize
   \par\noindent
   \footnotemark[$*$] Net exposure time.
   \par\noindent
   \footnotemark[$\dagger$] Source count rate of XIS (FI)/PIN for Suzaku and FPMA for
   NuSTAR in all energy bands.
   \par\noindent
   }\hss}
   }
  \end{tabular}
 \end{center}
\end{table*}

We used the archival data of $\gamma$ Cas with the Suzaku X-ray observatory
\citep{mitsuda07} and NuSTAR \citep{harrison13} and those of HD\,110432 with Suzaku
(table~\ref{t01}). No observation was made for HD\,110432 with NuSTAR. The Suzaku data
were presented previously in \citet{torrejon12,shrader15,hamaguchi16}, whereas the
NuSTAR data are presented here for the first time. We used the \texttt{HEADAS} software
package version 6.19 for the data reduction and the \texttt{Xspec} package version 12.9
for the spectral fitting throughout this paper.

\subsection{Suzaku}\label{s2-1}
Suzaku has two instruments in simultaneous operation: the X-ray Imaging Spectrometer
(XIS; \cite{koyama07}) and the Hard X-ray Detector (HXD;
\cite{kokubun07,takahashi07}). We use both instruments to cover a wide energy
range of 0.5--70~keV using the screened events through the Suzaku processing pipeline
version 2.7.16.30 with the standard set of event screening. The latest calibration
database (20160607 for XIS and 20110913 for HXD) were used.

\subsubsection{XIS}\label{s2-1-1}
We extracted the source events from a circle around the target with a 3\arcmin\ radius,
whereas the background events from the remaining part of the XIS sensors excluding a
6\arcmin\ radius circle around the target, a 3\arcmin radius circle around the onboard
$^{55}$Fe calibration sources, and field edges. $\gamma$ Cas was bright enough to cause
a slight photon pile-up \citep{yamada12}. We excluded the innermost circle with a
0\farcm07 radius in the source region and a 1\farcm7 width along the readout in the
background region.

\subsubsection{HXD-PIN}\label{s2-1-2}
We retrieved the latest catalogs by the INTEGRAL IBIS \citep{bird10} and Swift BAT
\citep{baumgartner13} instruments and confirmed that there are no contaminating sources
within the FWZI field of the two observations. We used Non X-ray background (NXB)
spectrum version 2.0 distributed by the instrument team \citep{fukazawa09}. The cosmic X-ray
background (CXB) spectrum, which is another major source of background, was generated by
convolving a spectral model \citep{boldt87} with the detector's spatial and spectral
responses. The NXB and CXB spectra were then merged and were subtracted as the PIN
background spectrum.

\subsection{NuSTAR}\label{s2-2}
NuSTAR has two co-aligned X-ray telescopes. At the focal plane of each telescope, a
Focal Plane Module (FPM) is placed, which is called FPMA or FPMB. We reprocessed the
$\gamma$ Cas data using the NuSTAR Data Analysis Software (\texttt{NuSTARDAS} v.1.4.1)
and the calibration files (2016 October 21) to obtain a cleaned event list. Source
events were extracted from a circle of 3\arcmin\ radius, while background events were
extracted from an annular region with an inner and outer radius of 4\arcmin and
6\arcmin.

\section{Analysis}\label{s3}
\begin{figure*}[htbp]
 \begin{center}
  \includegraphics[width=0.48\textwidth, bb=0 0 842 595, clip]{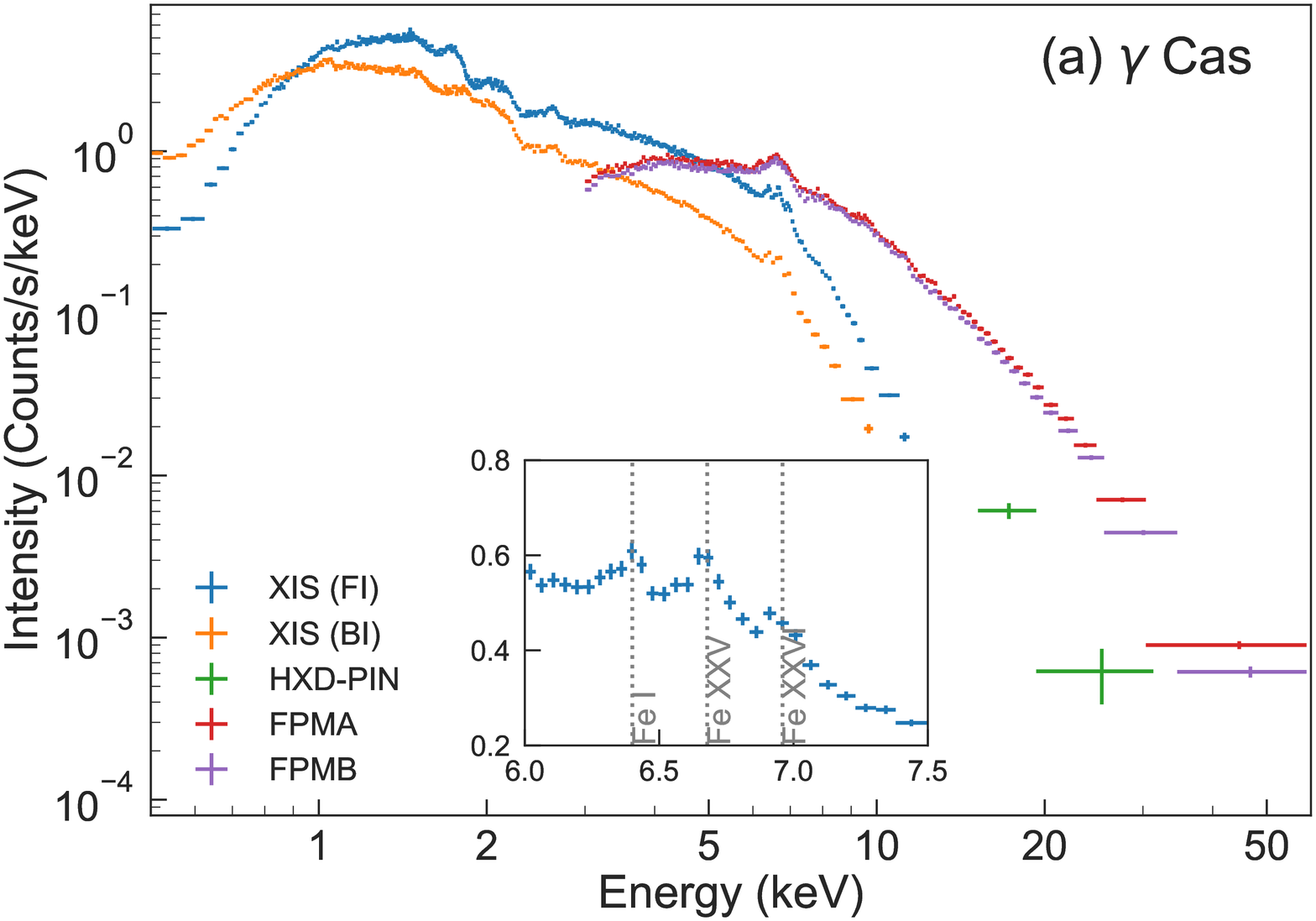}
  \includegraphics[width=0.48\textwidth, bb=0 0 842 595, clip]{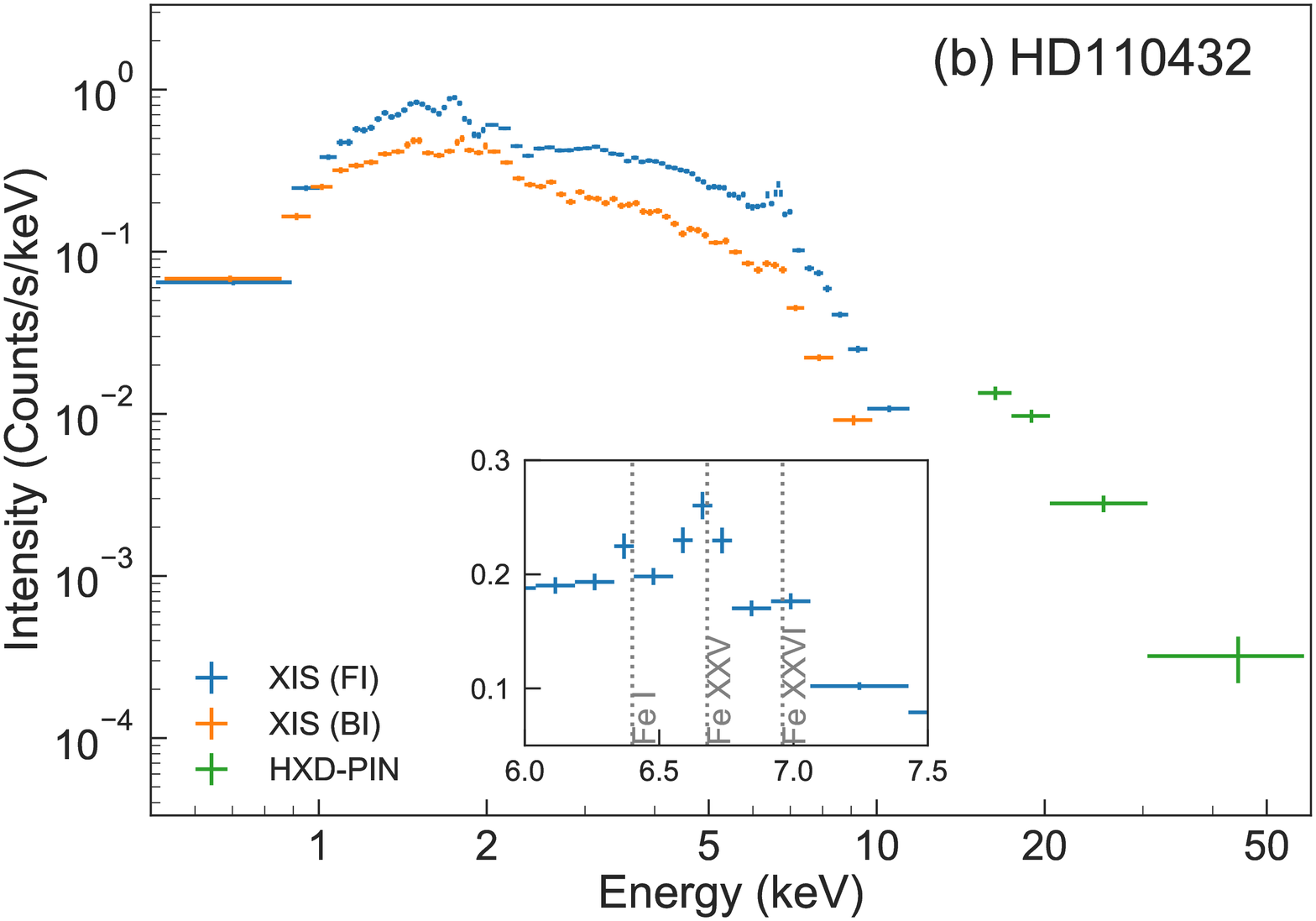}
 \end{center}
 \caption{Background-subtracted spectrum of (a) $\gamma$\,Cas and (b) HD\,110432. The
 Suzaku XIS FI, BI, HXD-PIN, and NuSTAR FMPA, FPMB spectra are shown with different
 colors. An enlarged view of the XIS FI spectrum in the Fe K band is shown in the inset.}
 \label{f02}
\end{figure*}

Figure~\ref{f02} shows the Suzaku and NuSTAR spectra of $\gamma$\,Cas and the Suzaku
spectra of HD\,110432. Both spectra are characterized by a very hard continuum extending
beyond 50~keV and spectral features of abundant species. The insets give a close-up view
of the Fe K-band, in which the prominent emission lines are found at 6.4, 6.7, and
7.0~keV respectively from Fe\emissiontype{I}, Fe\emissiontype{XXV}, and
Fe\emissiontype{XXVI}. The emission lines of highly-ionized Fe and other species
indicate the presence of thermal plasma of different temperatures, whereas the 6.4~keV
line indicates the presence of the reprocessed emission, which presumably occurs at the
WD surface. This should accompany a hard Compton reflection continuum, which should be
modeled consistently.

For the spectral analysis, we generated the XIS redistribution matrix functions and the
XRT ancillary response files using the \texttt{xisrmfgen} and \texttt{xissimarfgen}
tools, respectively \citep{ishisaki07}. The spectra by the two FI sensors were merged
for their nearly identical responses, while the BI spectrum was treated separately. We
used the entire energy band except for the 1.8--2.0~keV band to avoid the known
calibration inaccuracy due to the Si K edge. For the FPMs, we generated the
redistribution matrix function and the ancillary response files using the
\texttt{nuproducts} tool. Spectra were treated separately for the two units. The
normalization relative to XIS (FI) was fixed to the value given by the instrument
team\footnote{See
http://www.astro.isas.ac.jp/suzaku/doc/suzakumemo/suzakumemo-2008-06.eps.}  for XIS (BI)
and HXD, while it was treated as a free parameter for the FPMs.

\begin{figure*}[htbp]
 \begin{center}
  \includegraphics[width=1.0\textwidth]{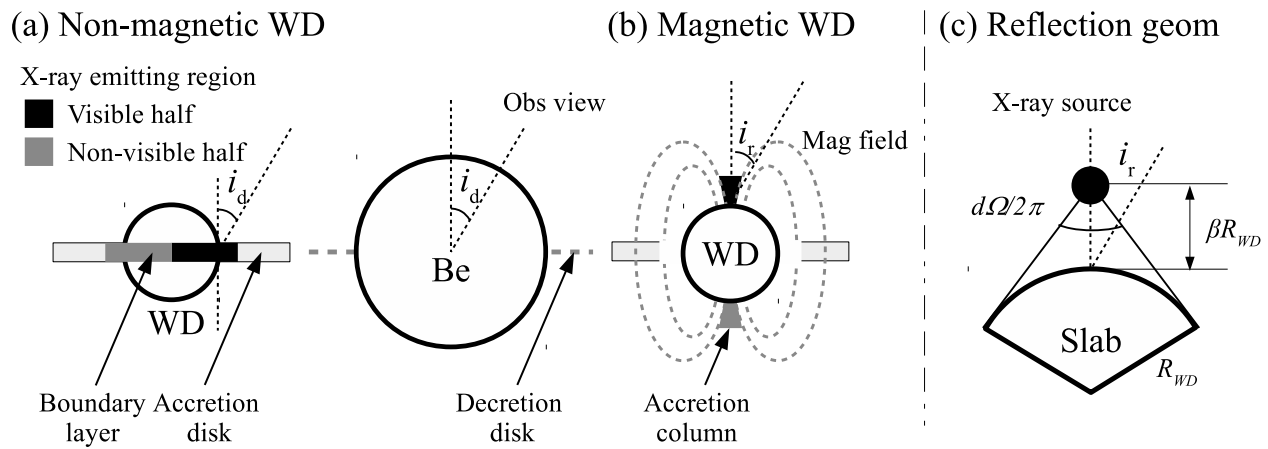}
 \end{center}
 \caption{Schematic view of the geometry: (a) non-magnetic WD with the boundary layer as
 the X-ray emitting site and (b) magnetic WD with the accretion column as
 the X-ray emitting site. A half of the X-ray emitting site is assumed visible from the
 observer. In (a), the WD accretion disk and the rotation axis is assumed to align with the
 B\textit{e} star decretion disk with a viewing angle of $i_{\mathrm{d}}$. (c) The geometry
 of the reflection modeling with a viewing angle of $i_{\mathrm{r}}$. The WD surface
 with a radius $R_{\mathrm{WD}}$ subtends an angle $d\Omega$/2$\pi$ seen from a
 point-like X-ray source at a height of $\beta R_{\mathrm{WD}}$ from the surface, in
 which $d\Omega/2\pi = 1- \sqrt{1-(\frac{1}{\beta+1})^{2}}$.}
 \label{f07}
\end{figure*}

\subsection{Non-magnetic accreting WD model}\label{s3-1}
\begin{figure*}[htbp]
 \begin{center}
  \includegraphics[width=0.48\textwidth, bb=0 0 842 595, clip]{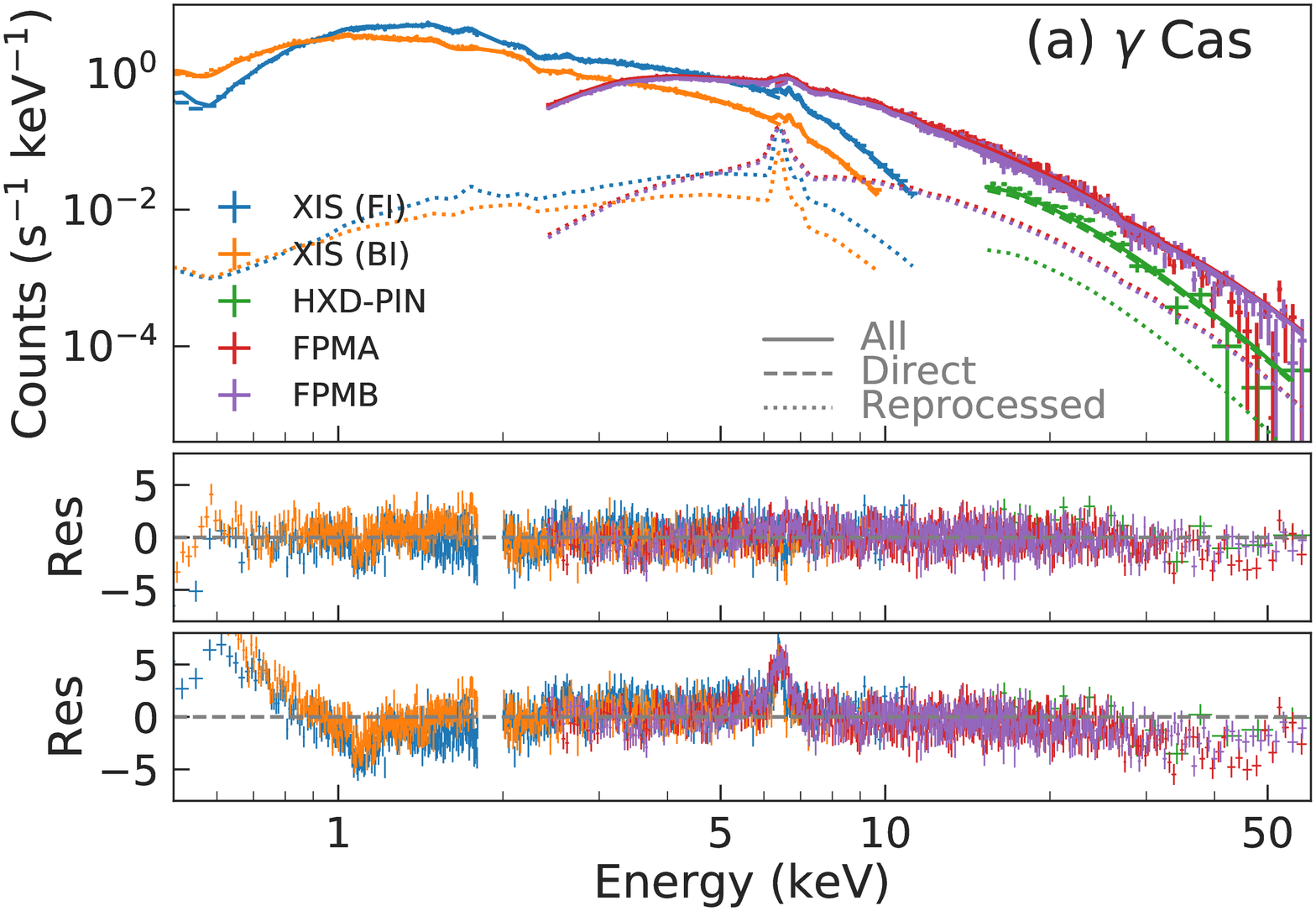}
  \includegraphics[width=0.48\textwidth, bb=0 0 842 595, clip]{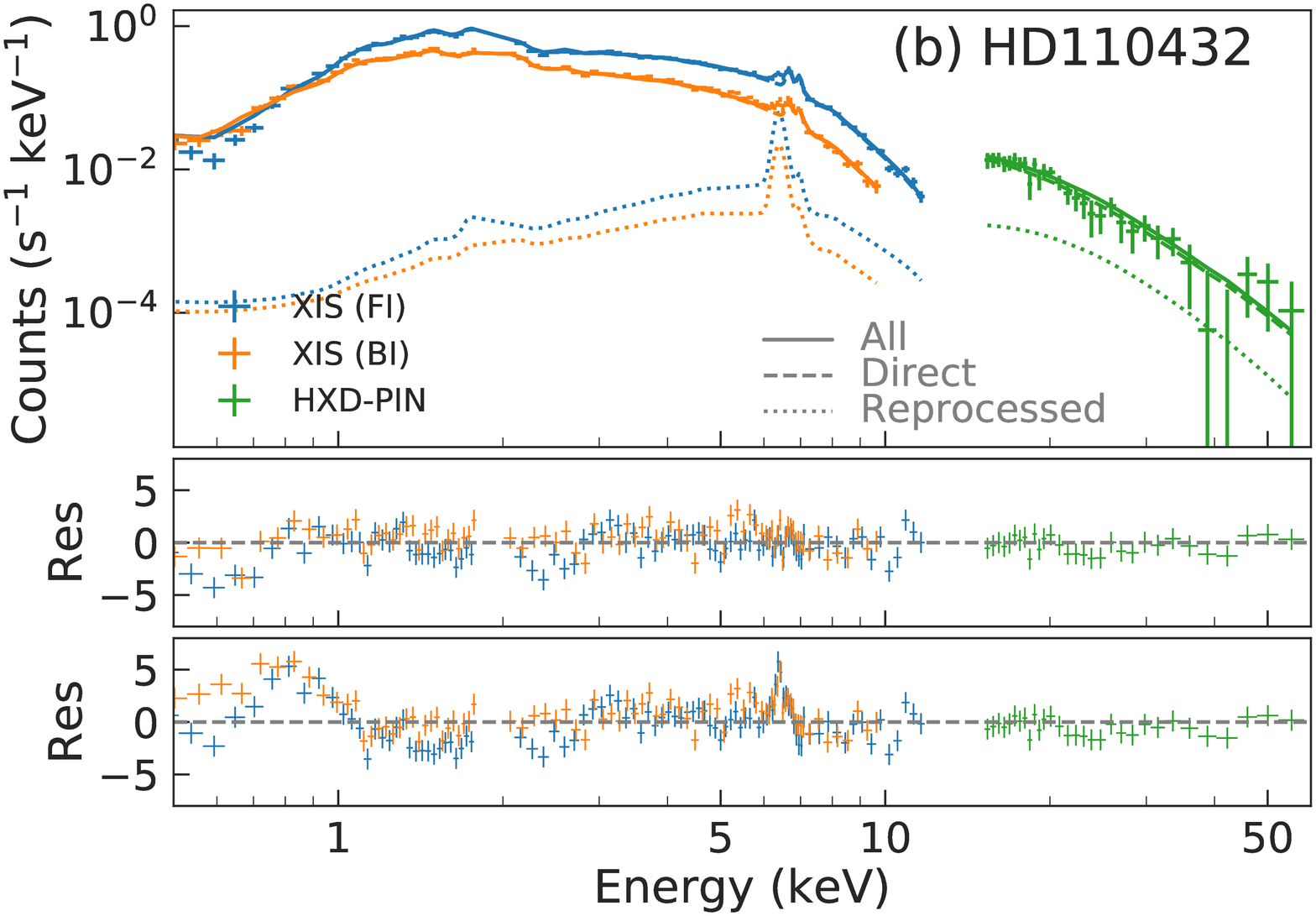}
 \end{center}
 \caption{Spectrum and the best-fit non-magnetic accreting WD model of (a) $\gamma$\,Cas
 and (b) HD\,110432. The upper panels show the data (plus marks) and the model (dotted,
 dashed, and solid curves respectively for the direct, reprocessed, and total
 components), whereas the middle panels show the residual to the fit. The bottom panel
 shows the residual to the fit by the simple model of only the cooling flow and the ISM
 extinction (\S~\ref{s3-1-2}).}
\label{f04}
\end{figure*}

\begin{table}[htbp]
 \caption{Best-fit parameters with the non-magnetic accreting WD model$^{*}$.}
 \label{t03}
 \begin{center}
  \begin{tabular}{llcc}
   \hline
   \hline
   Model            & Parameter & $\gamma$\,Cas & HD\,110432\\
   \hline
   \multicolumn{3}{l}{(Fixed values)}\\
   Distance$^{\dagger}$ & $D$ (pc)& 188 & 420\\
   Angle            & $i_{\mathrm{r}}$ (degree) & 65 & 59 \\
   \texttt{tbabs}$^{\ddagger}$ & $N_{\rm {H}}$ ($10^{20}$ cm$^{-2}$) & 1.45 & 15.8 \\
   \hline
   \multicolumn{3}{l}{(Fitted values$^{*}$)}\\
   \texttt{tbpcf}   & $N_{\rm {H}}$ ($10^{22}$ cm$^{-2}$) & 0.94$_{-0.02}^{+0.01}$ & 1.36$_{-0.03}^{+0.05}$\\
                    & Covering fraction & 0.486$\pm$0.003 & 0.893$\pm$0.005 \\
   \texttt{mkcflow} & $T_{\mathrm{max}}$ (keV) & 25.0$_{-0.12}^{+0.13}$& 48.2$_{-2.9}^{+1.7}$\\
                    & $Z$ (solar) & 0.41$_{-0.01}^{+0.01}$& 1.23$_{-0.10}^{+0.07}$\\
                    & $\dot{M}_{\mathrm{X}}$ (10$^{-10} M_{\odot}$~yr$^{-1}$) & 1.61$_{-0.01}^{+0.01}$& 1.82$_{-0.03}^{+0.14}$\\
   \texttt{reflect} & $d\Omega$/2$\pi$ & 0.27$_{-0.01}^{+0.01}$  &  0.52$_{-0.06}^{+0.05}$\\
   \texttt{gauss}   & EW$_{\mathrm{Fe}}$ (eV)    & 54$_{-4}^{+3}$& 90$_{-5}^{+4}$\\
   \hline
   \multicolumn{2}{l}{$\chi^{2}_{\mathrm{red}}$ (d.o.f.)} & 1.27 (2313) & 1.52 (200)\\
   \hline
   \multicolumn{4}{l}{\parbox{80mm}{
   \footnotesize
   \par \noindent
   \footnotemark[$*$] The errors indicate a 1$\sigma$ statistical uncertainty. \\
   \footnotemark[$\dagger$] The distance is given by \citet{vanleeuwen07} for $\gamma$
   Cas and by the Gaia data release 2 \citep{GaiaCollaboration2016,GaiaCollaboration2018}
   for HD\,110432.\\
   \footnotemark[$\ddagger$] The ISM extinction is derived from the $E(B-V)$ measurement
   of the B\textit{e} star \citep{Jenkins2009}.\\
   }}
  \end{tabular}
 \end{center}
\end{table}

\subsubsection{Model}\label{s3-1-1}
We now apply a physical model first by assuming that the two sources have a non-magnetic
accreting WD. In classical WD binaries, they are called dwarf novae (DNe;
\cite{warner95}). The magnetic field ($B_{\mathrm{WD}} < 0.1$~MG) is not strong enough
to disturb the mass accretion through the disk. In such systems, the accreting matter
rotating at a Keplerian velocity of the disk loses a large amount of kinetic energy when
it lands on the WD surface rotating at a much slower speed. The energy is converted to
heat and forms a plasma region (called the boundary layer) that expands from the equator
to higher latitudes on the WD surface with a temperature gradient
\citep{patterson85}. The emission is optically-thin for DNe at quiescence
\citep{pringle79}. During DN outbursts and super-outbursts, the X-ray spectrum deviates
from it \citep{wada17}. As the two sources are not known to exhibit such episodic
events, we do not discuss DNe during outbursts and super-outbursts. Similar, but less
well-studied systems are nova-like variables. They are considered to be at outbursts all
the time with a high accretion rate except for the decline in brightness in some
sources. The X-ray emission from them in the steady state is also considered from the
boundary layer \citep{pratt04}.

The X-ray spectrum from the boundary layer is well described by the cooling flow model,
which is a convolution of the thin-thermal plasma model with a 1D stack of plasma sheets
with temperatures changing isobarically. As the name suggests, the model was originally
developed for the gas in the cluster of galaxies \citep{mushotzky88}, which is now
considered too simplistic to be relevant \citep{peterson01}. However, the model turned
out to be very useful to describe the emission from the boundary layer of DNe, which is
shown in many sources \citep{pandel05,wada17}.

\subsubsection{Fitting}\label{s3-1-2}
\paragraph{Continuum}\label{s3-1-2-1}
We started with the \texttt{mkcflow} model in the \texttt{xspec} package for the cooling
flow emission. The free parameters are the maximum plasma temperature
($T_\mathrm{max}$), the metal abundance ($Z$) relative to the ISM value \citep{wilms00}, and the
mass accretion rate ($\dot{M}_{\mathrm{X}}$). The mass accretion rate assessed by the
X-ray spectroscopy gives a lower limit of the mass transfer rate from the donor
($\dot{M}$) as only a part is dissipated radiatively in the X-ray emitting region and a
part of it is visible. The minimum plasma temperature was fixed at the lowest allowed
value in the model (81~eV).

We attenuated the model with the extinction by the ISM using the \texttt{tbabs} model
\citep{wilms00}. This simple model failed to fit the data with three major features in
the residuals (the bottom panels in figure~\ref{f04}) for both sources: (i) a large
broad deviation in the soft band below $\sim$1~keV, (ii) a local excess at 6.4~keV, and
(iii) a broad curvature beyond 20~keV.

\paragraph{Absorption}\label{s3-1-2-2}
We thus attempt to improve the residual feature (i) first. This structure is typical for
spectra that are partially absorbed. In fact, a partial absorption is seen in some DNe
\citep{wada17}. We separated the absorption into two: that by the ISM and that by the
local material. The former was modeled by the full-covering neutral absorber model
(\texttt{tbabs}) with the absorption column fixed at the ISM value derived from the
$E(B-V)$ measurement of the B\textit{e} star \citep{Jenkins2009}. The latter was modeled
by the partially-covering neutral absorber model (\texttt{tbpcf}), in which the
absorption column and the covering fraction are the free parameters. This significantly
improved the fitting. We also tried the partially-covering ionized absorber model
(\texttt{zxipcf}) or the two \texttt{tbpcf} models with a different set of
parameters. They were not required by the data. As all of them yielded the similar
parameters for the continuum model, we adopted the simplest one with one \texttt{tbpcf}
model for the local absorption.

\paragraph{Reflection}\label{s3-1-2-3}
We next attempt to improve the residual features (ii) and (iii). They are naturally
interpreted respectively as the Fe fluorescence and the Compton hump due to the emission
reprocessed at the cold matter in the vicinity of the boundary layer, which is
presumably the WD surface. We assumed the reflection geometry as shown in
figure~\ref{f07} (c). We modeled the former with the \texttt{gauss} model and the latter
with the \texttt{reflect} model \citep{magdziarz95} independently and iterated the
fitting until the two yielded a consistent subtended angle $d\Omega/2\pi$.

To fix the viewing angle ($i_{\mathrm{r}}$), we made the following assumptions as shown
in figure~\ref{f07} (a): (i) the orientation is aligned between the B\textit{e}
decretion disk and the WD accretion disk, (ii) a half of the boundary layer along the WD
equator is visible, and (iii) the reflection occurs only at the WD surface. The viewing
angle of the reflection ($i_{\mathrm{r}}$) averaged over the visible half is related to
that of the disk ($i_{\mathrm{d}}$) by $\cos{i_{\mathrm{r}}} = 2/\pi
\sin{i_{\mathrm{d}}}$. Here, we assumed that the boundary layer has a negligible spread
in the latitude direction of the WD. Based on the optical and near-infrared observations
that spatially resolved the decretion disk, $i_{\mathrm{d}}=$~42 and 55 degrees
\citep{stee12,stee13}, thus $i_{\mathrm{r}}=$ 65 and 59 degrees respectively for
$\gamma$ Cas and HD\,110432.

\subsubsection{Result}\label{s3-1-3}
These modifications improved the fitting significantly as shown in the middle panels of
figure~\ref{f04}. We still see a residual structure below 0.6~keV and
$\sim$1.1~keV. These show the limitations of our model due to the too simplistic
assumptions of the cooling flow model to describe the boundary layer, not including the
soft ($\sim$0.5~keV) thermal X-ray emission due to the internal shocks of the
radiatively-driven wind commonly seen in early-type stars \citep{Berghofer1997a}, too
simplistic assumption of the geometry (figure~\ref{f07}), etc. We obtained some further
improvements by e.g, adding a soft thermal emission attenuated only by the ISM
extinction, but not much. These tweaks do not affect the best-fit parameters of the
continuum model significantly, as they are derived based on the spectrum above
$\approx$3~keV where the effect of the inaccurate modeling in the soft band is
limited. We, therefore, use this model and ignore the data below 0.6~keV to derive the
best-fit values in table~\ref{t03}.

\subsection{Magnetic accreting WD model}\label{s3-2}
\begin{figure*}[htbp]
 \begin{center}
  \includegraphics[width=0.48\textwidth, bb=0 0 842 595, clip]{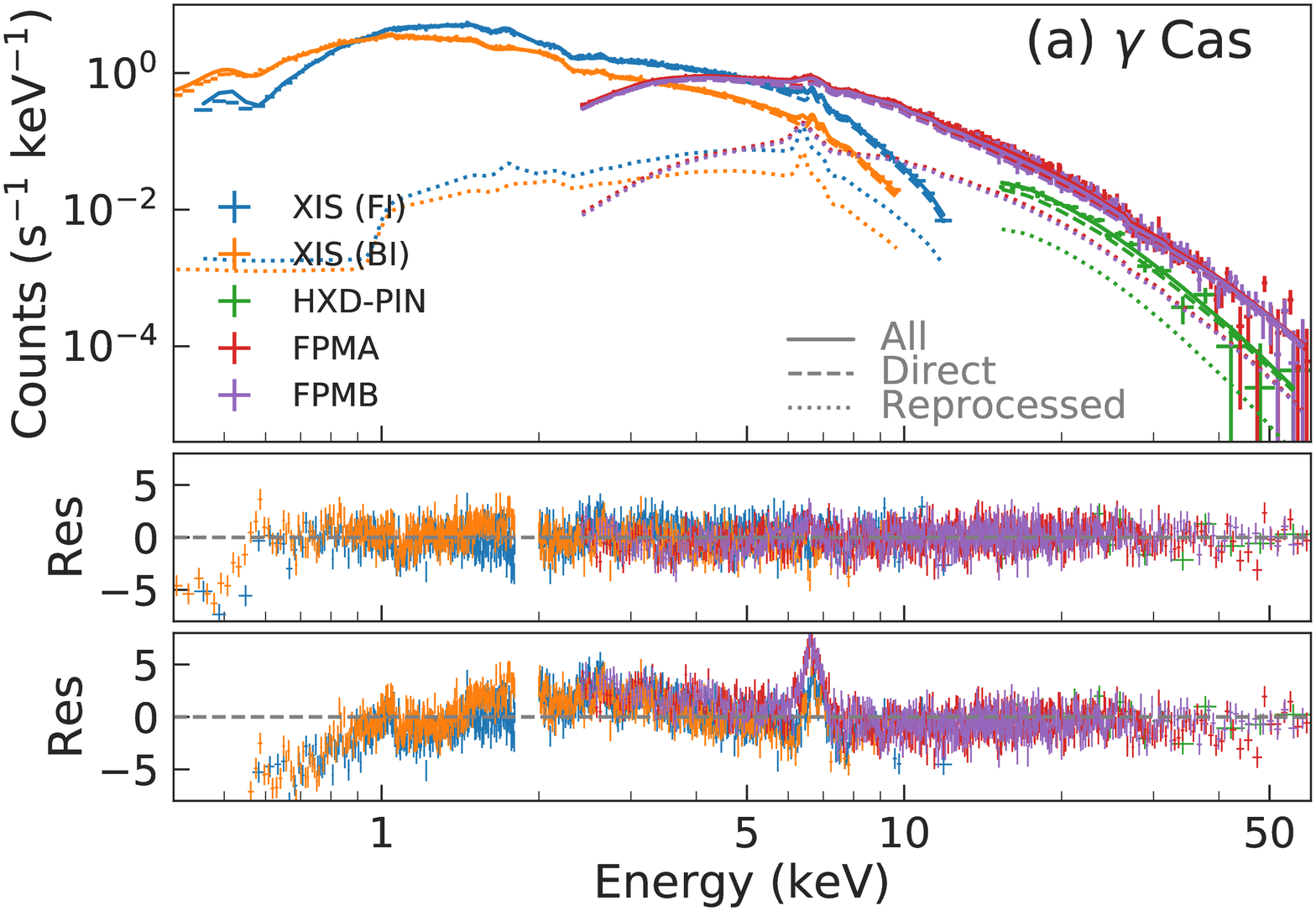}
  \includegraphics[width=0.48\textwidth, bb=0 0 842 595, clip]{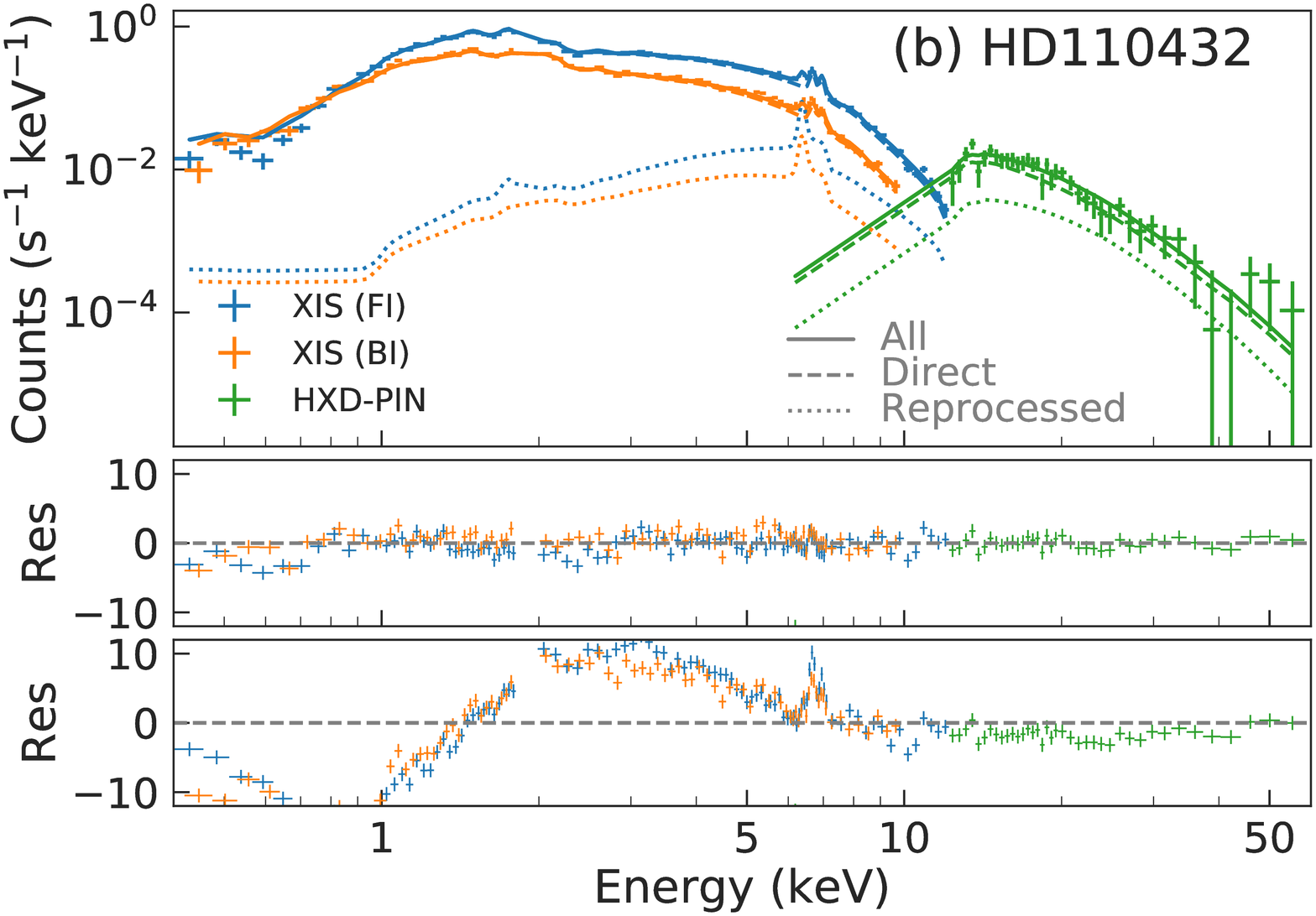}
 \end{center}
 \caption{Spectrum and the best-fit magnetic accreting WD model of (a) $\gamma$\,Cas and
 (b) HD\,110432. The symbols follow figure~\ref{f04}. The Compton reflection is modeled
 only above 1~keV, below which the component is negligible.}
 \label{f05}
\end{figure*}

\begin{table}[htbp]
 \caption{Best-fit parameters with the magnetic accreting WD model$^{*}$.}
 \label{t04}
 \begin{center}
  \begin{tabular}{llcc}
   \hline
   \hline
   Model            & Parameter & $\gamma$\,Cas & HD\,110432\\
   \hline
   \multicolumn{3}{l}{(Fixed values)}\\
   Distance$^{\dagger}$ & $D$ (pc)& 188 & 420 \\
   \texttt{tbabs}$^{\ddagger}$ & $N_{\rm {H}}$ ($10^{20}$ cm$^{-2}$) & 1.45 & 15.8 \\
   \hline
   \multicolumn{3}{l}{(Fitted values$^{*}$)}\\
   \texttt{tbpcf}   & $N_{\rm {H}}$ ($10^{22}$ cm$^{-2}$) & $0.66_{-0.01}^{+0.02}$ & $0.94_{-0.02}^{+0.03}$\\
                    & Covering fraction & 0.453$_{-0.004}^{+0.004}$ & 0.865 $_{-0,006}^{+0.006}$ \\
   \texttt{acrad}   & $M_{\mathrm{WD}}$ ($M_{\odot}$) & $0.72_{-0.06}^{+0.04}$ & $0.81_{-0.03}^{+0.03}$ \\
                    & $Z$ (solar$^{\S}$) & $0.37 \pm 0.01$ & $0.92 \pm 0.07$ \\
                    & $\log{a}$ (g~cm$^{-2}$~s$^{-1}$) & $-0.61_{-0.11}^{+0.35}$ & $3.69_{-3.8}^{+\infty}$ \\
                    & $i_{\mathrm{r}}$ (degree) & $68.5_{-1.1}^{+1.2}$ & $69.2_{-3.6}^{+3.3}$ \\
   \multicolumn{3}{l}{(Derived values$^{\|}$)}\\
                    & $L_{\mathrm{X}}$ (erg~s$^{-1}$) & 1.0$\times 10^{33}$ & 2.5$\times 10^{33}$ \\
                    & $\dot{M}_{\mathrm{X}}$ ($M_{\odot}$~yr$^{-1}$) & 1.3$\times 10^{-10}$  & 2.5$\times 10^{-10}$ \\
                    & $f$  & (4--6)$\times$10$^{-3}$ & $>$0 \\
   \hline
   \multicolumn{2}{l}{$\chi^{2}_{\mathrm{red}}$ (d.o.f.)} & 1.23 (2312) & 1.42 (212) \\
   \hline
   \multicolumn{4}{l}{\parbox{80mm}{
   \footnotesize
   \par \noindent
   \footnotemark[$*$] The errors indicate a 1$\sigma$ statistical uncertainty. \\
   \footnotemark[$\dagger$] The distance is given by \citet{vanleeuwen07} for $\gamma$
   Cas and by the Gaia data release 2 \citep{GaiaCollaboration2016,GaiaCollaboration2018}
   for HD\,110432.\\
   \footnotemark[$\ddagger$] The ISM extinction is derived from the $E(B-V)$ measurement
   of the B\textit{e} star \citep{Jenkins2009}.\\
   \footnotemark[$\S$] Assuming that Fe represents the metals, the difference of the
   Fe abundance between the \citet{anders89} and \citet{wilms00} is corrected to match
   with the latter.\\
   \footnotemark[$\|$] $L_{\mathrm{X}}$ is derived by integrating the best-fit
   \texttt{acrad} model in the 0.2--100~keV band. $\dot{M}_{\mathrm{X}}$ and $f$ are
   derived using equations~\ref{e01} and \ref{e02}.\\
   }}
  \end{tabular}
 \end{center}
\end{table}

\subsubsection{Model}\label{s3-2-1}
Next, we apply a physical model assuming that the two sources have a magnetic accreting
WDs. They are called either intermediate polars (IPs) for a moderate field (0.1--10~MG)
or polars for a strong field ($>$10~MG). In IPs, the accretion disk is truncated and the
accreting matter falls along the magnetic field. In polars, the accretion takes place
without intervening an accretion disk. Whichever the case, the strong shock is formed
above the magnetic poles on the WD surface. The kinematic energy is dissipated into heat
at the shock, which cools radiatively. The geometry is different from the boundary
layer, which makes some differences in the X-ray spectra in comparison to non-magnetic
accreting WDs. Many models have been developed to describe the X-ray emission from the
post-shock accretion column with successful applications to real data
\citep{wu94,cropper99,suleimanov05,yuasa10,hayashi14a,hayashi14b}.

X-ray spectra of magnetic accreting WDs are also characterized by the ubiquitous presence of a
partially covering local absorption
\citep{Norton1989,Patterson1994,Done1998,Ramsay2008}, which is considered to stem from
the absorption by the pre-shock accretion flow around the X-ray emitting post-shock
region.

Among the magnetic accreting WD binaries, polars are often characterized by the presence
of a very bright soft X-ray emission component \citep{ishida97}, which is not seen in
the two $\gamma$ Cas analogs. However, the absence of this component alone does not
argue against polars, as some polars do not exhibit such component \citep{Ramsay2004}.

\subsubsection{Fitting}\label{s3-2-2}
We use our model (\texttt{acrad}; \cite{hayashi14a}), which calculates the X-ray
emission from the post-shock accretion column of magnetic accreting WDs. This model also
calculates the Compton and fluorescence emission consistently \citep{hayashi18} and fits
both the direct and reprocessed components simultaneously in a geometry shown in
figure~\ref{f07} (b). We note, however, that the cyclotron cooling is not included in
the model, hence the result is subject to some systematic effects \citep{wu94,cropper99}
if the two sources host a strongly magnetic WD.

The free parameters of the model are (i) the WD mass ($M_\mathrm{WD}$), (ii) the
specific mass accretion rate, or the mass accretion rate per unit area
($a=\dot{M}_{\mathrm{X}}/S_{\mathrm{col}}$), in which $S_{\mathrm{col}}$ is the cross
section at the foot of the accretion column, (iii) the metal abundance ($Z$) relative to
solar \citep{anders89}, and (iv) the viewing angle of the reflection
$i_{\mathrm{r}}$. Based on the best-fit parameters and the following relations:
\begin{equation}
 L_{\mathrm{X}} = \frac{GM_{\mathrm{WD}}\dot{M}_{\mathrm{X}}}{R_{\mathrm{WD}}}\label{e01}
\end{equation}
\begin{equation}
 f = \frac{S_{\mathrm{col}}}{4\pi R_{\mathrm{WD}}^{2}},\label{e02}
\end{equation}
in which $L_{\mathrm{X}}$ is the total luminosity of the plasma emission and $G$ is the
gravitational constant, we further derived $\dot{M}_{\mathrm{X}}$ and the fractional
area $f$ of the accretion column on the WD surface.

We started with the \texttt{acrad} model without the reprocessed components modified by
the ISM extinction using the \texttt{tbabs} model. As shown in the bottom panels in
figure~\ref{f05}, a large residual is found in both sources. This indicates the presence
of a partial absorber and the reprocessed component, thus we took the same approach with
the non-magnetic modeling (\S~\ref{s3-1-2-2}) to add a partial neutral covering model
and the reprocessed component. This improved the fitting significantly.  The result is
shown in figure~\ref{f05} and table~\ref{t04}.

\section{Discussion}\label{s4}
\subsection{Comparison to typical classical WD binaries}\label{s4-1}
\begin{figure*}[htbp]
 \begin{center}
 \includegraphics[width=0.48\textwidth, bb=0 0 595 595, clip]{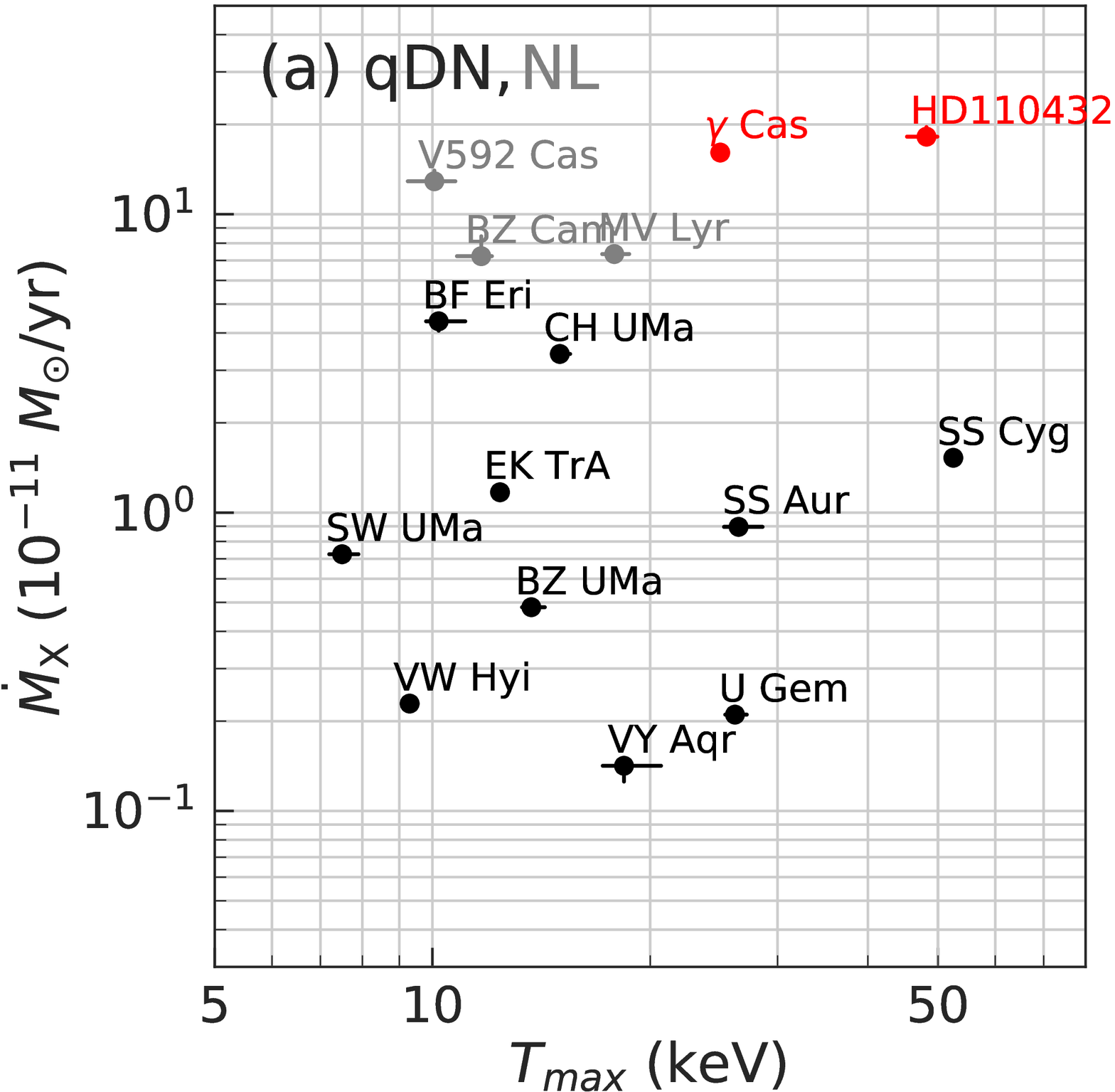}
 \includegraphics[width=0.48\textwidth, bb=0 0 595 595, clip]{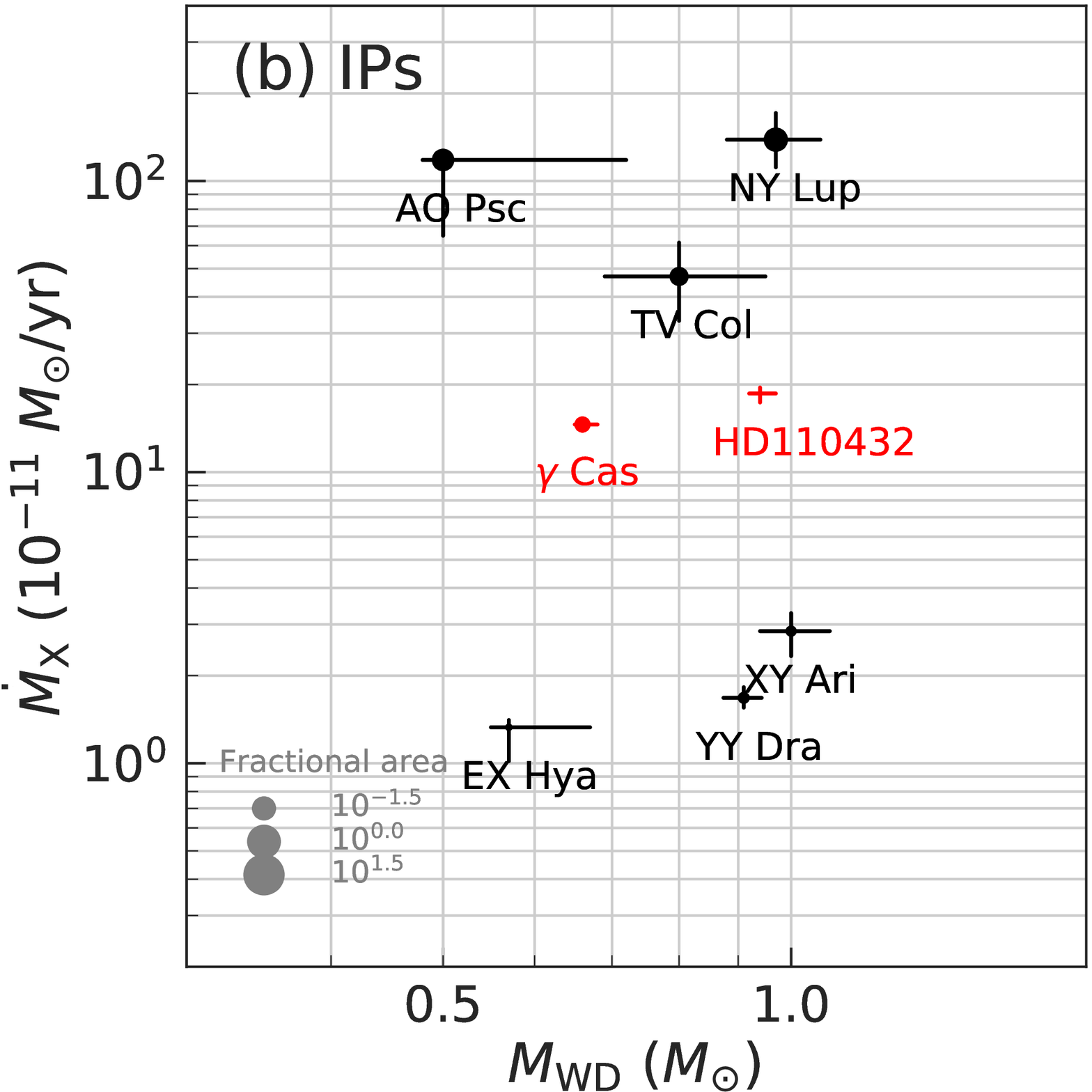}
 \end{center}
 \caption{Comparison between the $\gamma$ Cas analogs (red) and classical WD binaries:
 (a) DNe at quiescence (qDNe; black) and nova-like variables (NL; gray) and (b)
 Intermediate polars (IPs; black). In (b), the symbol size represents the fractional
 area of the accretion column $f$, which was not constrained for HD\,110432. The values
 for DNe are from \citet{wada17} using Suzaku XIS, those for NLs are from this work
 using Swift X-Ray Telescope and NuSTAR FPM, and those for IPs are also from this work
 using Suzaku XIS and HXD. The distance is based on the Gaia data release 2
 \citep{GaiaCollaboration2016,GaiaCollaboration2018} when available.}
\label{f06}
\end{figure*}

We applied the physical models of non-magnetic and magnetic accreting WDs for the two
$\gamma$ Cas analogs (\S~\ref{s3-1} and \S~\ref{s3-2}, respectively). We now compare the
result with that of their typical counterparts in classical WD binaries. We argue that
the two $\gamma$ Cas analogs can be interpreted reasonably by both of the models
within a parameter range of $M_{\mathrm{WD}}$ and $\dot{M}_{\mathrm{X}}$ obtained for
the classical WD binaries.

\begin{figure}[htbp]
\begin{center}
 \includegraphics[width=0.48\textwidth, bb=0 0 595 595, clip]{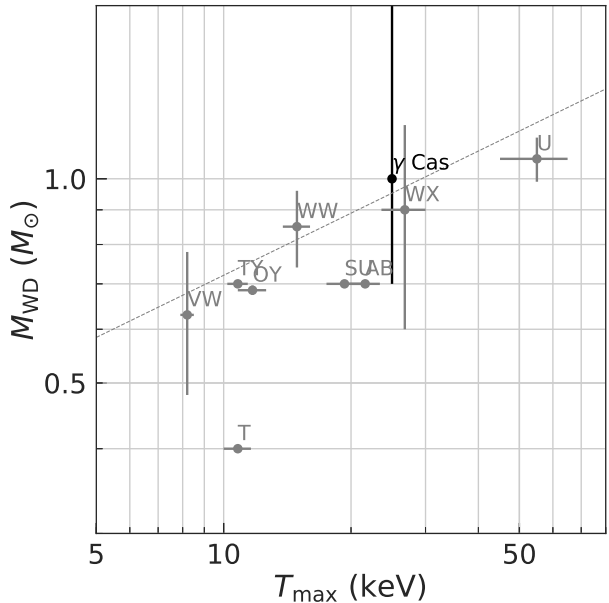}
 \end{center}
 \caption{Scatter plot of DNe at quiescence between $M_{\mathrm{WD}}$ by optical
 spectroscopy versus $T_{\mathrm{max}}$ by X-ray spectroscopy. DNe data are from
 \citet{pandel05} for T Leo, OY Car, VW Hyi, WX Hyi, SU UMa, TY, PsA, AB Dra, WW Cet,
 and U Gem. The optical data of $\gamma$ Cas are from \citet{harmanec00}. The linear
 regression line is shown with the dotted line.}
 \label{f10}
\end{figure}

\subsubsection{Non-magnetic case}\label{s4-1-1}
For comparison, we used the result by \citet{wada17} who compiled the X-ray spectra of
14 DNe at quiescence (qDNe) using the archival Suzaku data and fitted them with the
cooling flow model. We plotted $T_{\mathrm{max}}$ and $\dot{M}_{\mathrm{X}}$ in
figure~\ref{f06} (a).

On the one hand, the $T_{\mathrm{max}}$ values of the two $\gamma$ Cas analogs are
comparable to those of qDNe, indicating that the WD mass is in a reasonable range. A
relation is known between $T_{\mathrm{max}}$ and $M_{\mathrm{WD}}$ for qDNe with an
optical $M_{\mathrm{WD}}$ estimate \citep{Byckling2010}. We approximated their physical
relation with a linear regression, based on which we estimated $M_{\mathrm{WD}}$ from
$T_{\mathrm{max}}$ (figure~\ref{f10}). $\gamma$ Cas is placed along the linear
regression line with its optically-derived compact star mass of 0.7--1.9 $M_{\odot}$
\citep{harmanec00}.

On the other hand, the $\dot{M}_{\mathrm{X}}$ values of the two $\gamma$ Cas analogs are
significantly higher than those of qDNe. This is still the case when we compare with the
nova-like variables; we plotted results by applying the same model for three sources
using the Swift and NuSTAR data in figure~\ref{f06} (a). In fact, $\dot{M}_{\mathrm{X}}$
values of nova-like variables are not very high despite the fact that the $\dot{M}$
estimates by optical and ultra-violet (UV) are much higher than those of qDNe
\citep{balman14}. However, this alone does not argue against the possibility that the
two $\gamma$ Cas analogs host non-magnetic accreting WDs. The optically-thin boundary
layer may be maintained at a much higher mass accretion rate of
$\approx$10$^{-9}~M_{\odot}$~yr$^{-1}$\citep{Luna2018} if the form of accretion is
different. The two $\gamma$ Cas analogs may be such cases.

\subsubsection{Magnetic case}\label{s4-1-2}
For the magnetic case, we applied the same model described in \S~\ref{s3-2} to several
selected IPs using the Suzaku data for comparison. We plotted $M_{\mathrm{WD}}$ and
$\dot{M}_{\mathrm{X}}$ in figure~\ref{f06} (b). Both $\gamma$ Cas and HD\,110432 are
comparable to known IPs in both parameters. The validity of the $M_{\mathrm{WD}}$
estimates based on the X-ray spectral modeling is confirmed in several studies
\citep{suleimanov05,yuasa10}.  The $f$ value (eqn.~\ref{e02}), which is expressed by the size of the
symbol in figure~\ref{f06} (b), is well constrained for $\gamma$ Cas with a small value
of $\lesssim$1\%. Such a small value is common in IPs, and is confirmed in eclipsing
systems \citep{hellier97}. The X-ray estimate of $M_{\mathrm{WD}}$ for $\gamma$ Cas is
again consistent with the optically-derived one.

\subsection{Comparison between non-magnetic and magnetic cases}\label{s4-2}
We examine additional pieces of evidence to discriminate between the non-magnetic and
magnetic cases. There are several spectral features that appear differently between
these two for classical WD binaries. We argue, here, that the two $\gamma$ Cas analogs
cannot be classified conclusively to either one of them because the two sources exhibit
mixed features of the two cases and these features have overlapping distributions in
classical WD binaries.

\subsubsection{Partial covering}\label{s4-2-1}
The partial covering absorber is the first feature to examine. It is seen in the majority of
IPs \citep{Ramsay2008}. In contrast, DNe do not show evidence of partial absorption in
general \citep{wada17}, but some DNe do especially those with high inclination angles
(e.g., V893 Sco and OY Car; \cite{mukai09,pandel05}). In such systems, a part of the
boundary layer emission can be partially covered by the inner part of the accretion
disk.

\begin{figure}[htbp]
 \begin{center}
  \includegraphics[width=0.9\columnwidth, bb=0 0 583 566]{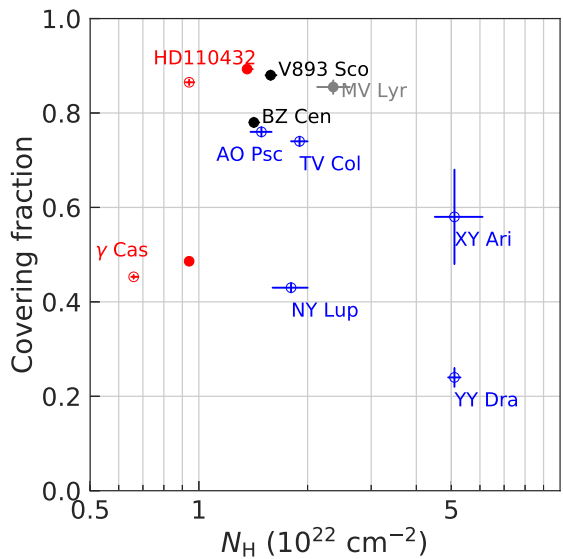}
 \end{center}
 \caption{Scatter plot of partial covering absorber. The two $\gamma$ Cas analogs (red),
 one nova-like variable (filled gray), DNe at quiescence (filled black; \cite{wada17}),
 and IPs (open blue), in which a partial covering absorber is required in the
 fitting. For the two $\gamma$ Cas analogs, the result by the non-magnetic model (open;
 table~\ref{t03}) and that by the magnetic model (filled; table~\ref{t04}) are
 separately shown.}
 \label{f12}
\end{figure}

The amount of partial covering column is subject to a large systematic uncertainty
depending on the continuum model to be applied. In order to avoid this, we employed the
same model for the comparison sources in figure~\ref{f06} for the amount of partial
covering column and its covering fraction. Figure~\ref{f12} shows a scatter plot to
compare the two $\gamma$ Cas analogs and DNe requiring a partial absorption \citep{wada17}, one
nova-like variable among three that requires a partial absorption, and several IPs. Some
sources required double partial covering, for which we plotted the one with a larger
covering fraction. The $\gamma$ Cas has a small absorption column for an IP, which may
argue against the magnetic case. Both $\gamma$ Cas and HD\,110432 exhibit a column
smaller than the two DNe in the comparison set, but they are distinctively different from
other DNe not requiring the partial absorption. Given the poor characterization of the
comparison sources and the mixed feature of the two $\gamma$ Cas analogs, it is
difficult to conclude for either of the two cases.

\subsubsection{Fe L lines}\label{s4-2-2}
\begin{figure*}[htbp]
 \begin{center}
  \includegraphics[width=0.9\textwidth, bb=0 0 595 842, clip]{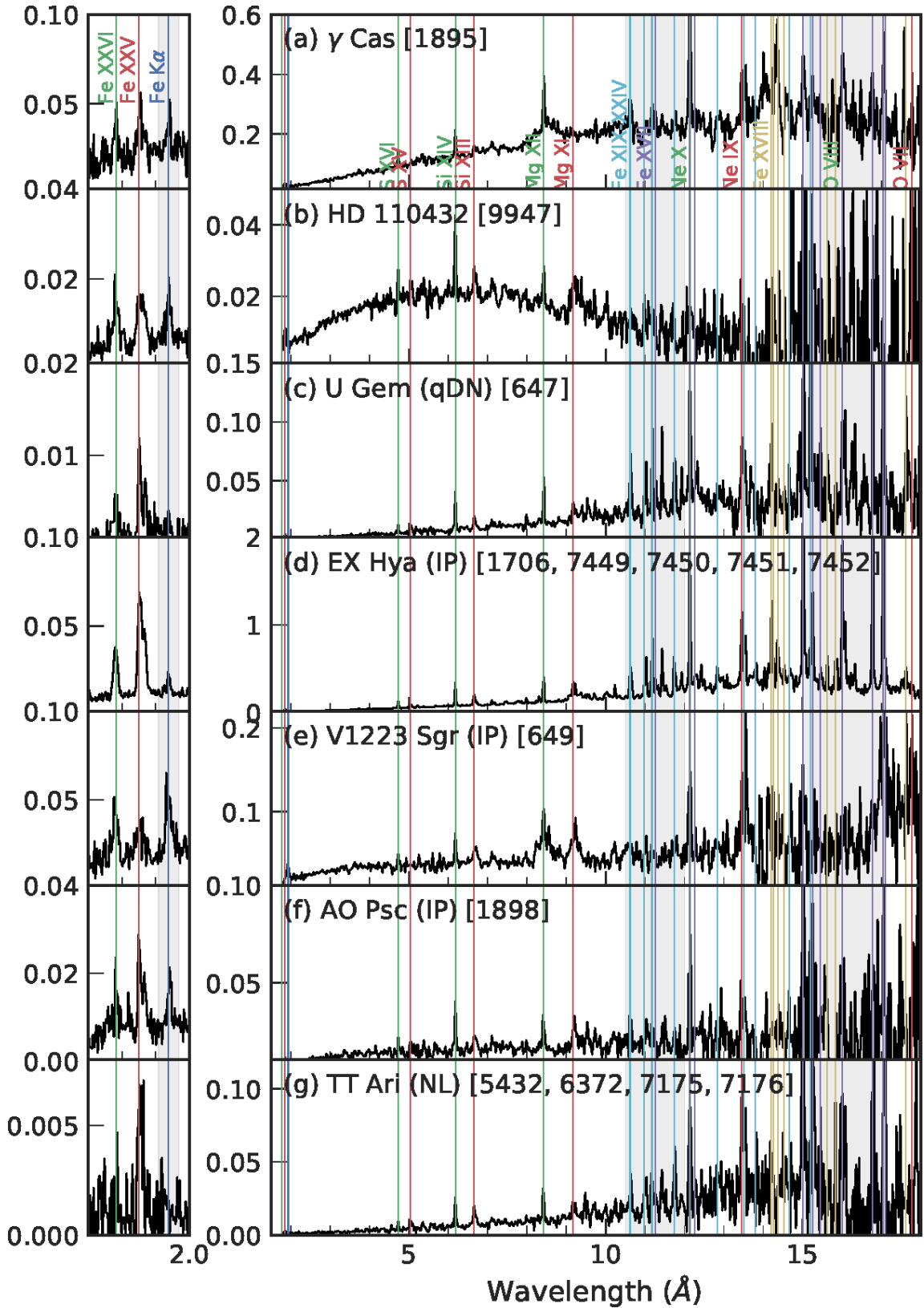}
 \end{center}
 \caption{Chandra HEG (left) or MEG (right) $\pm$1 order spectra combined respectively
 for 1.7--2.0 or 1.8--20~\AA\ of the two $\gamma$ Cas analogs, a DN at quiescence
 (qDN; U Gem), three IPs (EX Hya, V1223 Sgr, and AO Psc), and one nova-like variable
 (NL). The numbers in the parenthesis show the Chandra observation number. The ordinate
 shows the observed count rate divided by the effective area at the wavelength averaged
 over multiple observations. The gray shaded regions are the conspicuous differences
 between DNe and IPs for Fe \emissiontype{XIX--XIV} lines (in cyan) at 10--12 \AA, Fe
 \emissiontype{XVII} lines (in purple) at 15--17 \AA, and Fe K$\alpha$ line (in blue) at
 1.9 \AA.}
 \label{f08}
\end{figure*}

In the soft X-ray band, it is known that DNe at quiescence and IPs show a clear
dichotomy, though some exceptions, in high-resolution grating spectra
\citep{mukai03,pandel05}. The most prominent difference is found in the moderately
ionized Fe \emissiontype{XIX--XIV} lines at 10--12 \AA\ and Fe \emissiontype{XVII} lines
at 15--17 \AA\ as opposed to the highly ionized Fe \emissiontype{XXV--XXVI} lines at
1.7--2.0~\AA.

\citet{mukai03} argued that the moderately ionized lines should appear in the cooling
flow model as it is a synthesis of plasma emission of a wide range of temperatures. In
fact, the X-ray spectra of DNe at quiescence show these features
\citep{mukai03,pandel05}. Similar features are found in nova-like variables
\citep{zemko14}. On the other hand, the X-ray spectra of IPs in this wavelength range is
represented by a photo-ionized plasma model, thus the moderately ionized Fe lines are
weak in contrast to the highly ionized lines with an exception of EX Hya
\citep{mukai03}.

In figure~\ref{f08}, the archival Chandra high-energy transmission grating (HETG)
spectra are compared between the two $\gamma$ Cas analogs and some classical WD
binaries. The moderately ionized Fe lines are detected in $\gamma$ Cas
\citep{Smith2004}, though not as strong as non-magnetic WD samples shown in the
figure. HD\,110432 is noisy due to its larger extinction than $\gamma$ Cas. A more
qualitative assessment is needed in both classical WD binaries and the $\gamma$ Cas
analogs to use these features for classification.

\subsubsection{Fe K$\alpha$ line}\label{s4-2-3}
\begin{figure}[htbp]
 \begin{center}
  \includegraphics[width=0.9\columnwidth, bb=0 0 595 595, clip]{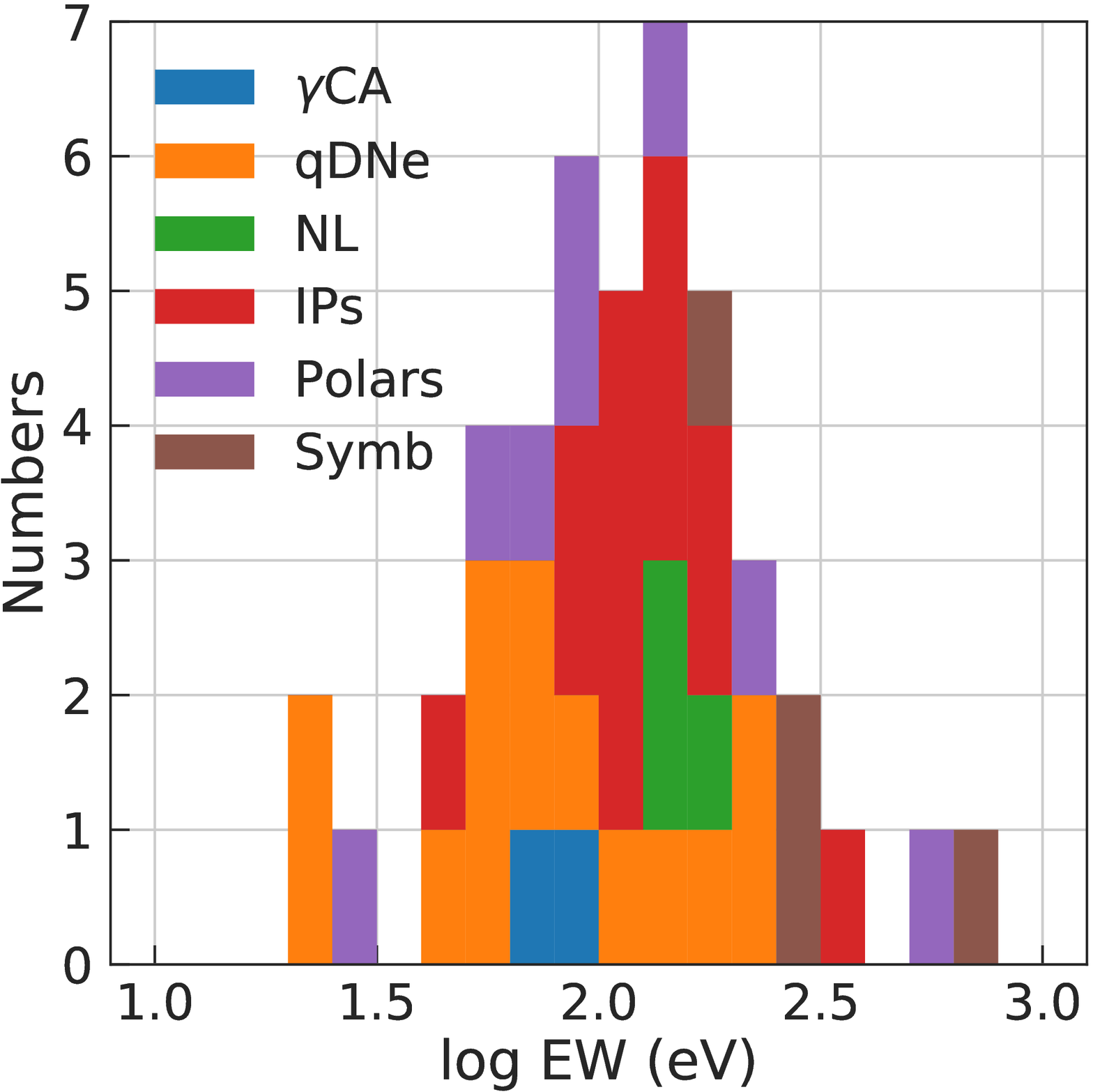}
 \end{center}
 \caption{Histogram of the EW of the Fe \emissiontype{I} fluorescence for the two
 $\gamma$ Cas analogs, DNe at quiescence \citep{wada17}, nova-like variables (NL; this
 work), IPs and polars \citep{ezuka99}, and hard symbiotic stars (Symb;
 \cite{Mukai2007,Smith2008,Kennea2009,Eze2010,Luna2018}).}
 \label{f13}
\end{figure}

Yet another is the equivalent width (EW) of the Fe \emissiontype{I} fluorescence. In
figure~\ref{f08} (left panels), the same data are compared in the 1.7--2.0~\AA\
range. IPs are known to have a stronger Fe fluorescence line than DNe and nova-like
variables in general. However, the distribution has a large overlap. In
figure~\ref{f13}, we plotted a stacked histogram of the EW of the two $\gamma$ Cas
analogs (table~\ref{t03}) and the comparison sources of the classical WD binaries. The
two $\gamma$ Cas analogs are distributed in the overlapped region of the non-magnetic
and magnetic cases, and a firm conclusion cannot be derived.

In the comparison, we added a small group of symbiotic stars with exceptionally hard
X-ray emission. Four such sources are known (RT Cru, T CrB, CD--57 3057, and CH Cyg;
\cite{Mukai2007,Smith2008,Kennea2009,Eze2010,Luna2018}). The origin of the hard X-ray
emission is debated whether it is from the boundary layer or not
\citep{Luna2018,Ducci2016}. These sources are characterized by an exceptionally large EW
of the Fe \emissiontype{I} fluorescence, suggesting that the reprocessed emission is
enhanced in a geometry different from the ones assumed in figure~\ref{f07}.  The two
$\gamma$ Cas analogs do not show such a large EW, thus they are not likely akin to the
hard symbiotic stars.



\subsection{Further observations}\label{s4-3}
We presented the X-ray spectral data in this paper and discussed that the two $\gamma$
Cas analogs can be interpreted as accreting WD binaries, though they are not
conclusively classified to either the non-magnetic or magnetic WD binaries. Finally, we
discuss some observations, especially long-term temporal behaviors in the X-rays, which
are important to elucidate the nature of the $\gamma$ Cas analogs further.

\subsubsection{X-ray pulsation}\label{s4-3-1}
A defining characteristic of an IP is a coherent X-ray pulsation due to the WD spin and
the localized X-ray emitting region \citep{mukai17}. For a similarly small fractional
area $f$ of the X-ray emitting region at least for $\gamma$ Cas when applied the
magnetic model (\S~\ref{s3-2}) , we would naturally expect a coherent pulsation in their
X-ray light curves. However, no such signal is confirmed to date. There are two
possibilities. One is that the magnetic pole is almost completely aligned to the
rotational axis (figure~\ref{f07}b), thus the flux change is too small to be
detected. Another is that the coherent signals are not investigated well in the expected
range of the periods.

\begin{figure}[htbp]
 \begin{center}
  \includegraphics[width=1.0\columnwidth, bb=0 0 667 595, clip]{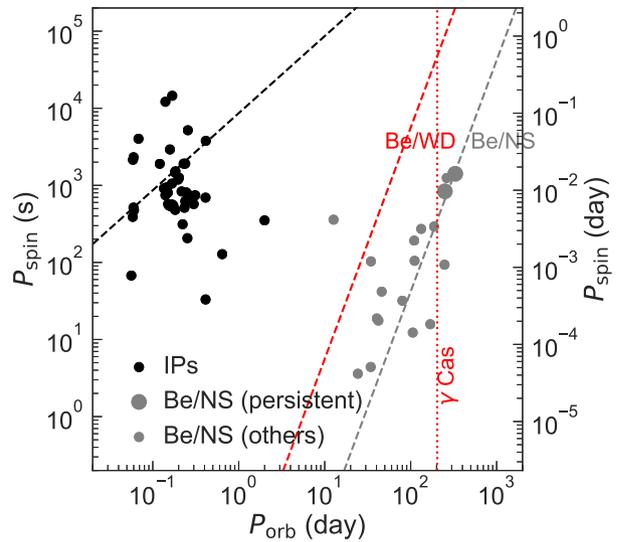}
 \end{center}
 \caption{Relations between the orbital period ($P_{\mathrm{orb}}$) and the spin period
 ($P_{\mathrm{spin}}$) for IPs (black; \cite{mukai17}) and B\textit{e}/NS binaries
 (gray; \cite{reig11}). For the latter, a larger symbol size is used for the persistent
 (X Per type) sources than the others. The phenomenological linear relation is shown with
 dashed lines for each class. The relation for B\textit{e}/WD binaries scaled from the
 one for the B\textit{e}/NS binaries \citep{apparao02} is shown with the red dashed
 line. The $P_{\mathrm{orb}}$ is shown for $\gamma$ Cas \citep{harmanec00} with the red
 dotted vertical line.}
 \label{f11}
\end{figure}

\citet{harmanec00,miroshnichenko02} discovered that $\gamma$ Cas is a binary system with
an orbital period of 202.59 days. Relations are known between the spin
($P_{\mathrm{spin}}$) and the orbital ($P_{\mathrm{orb}}$) periods for some established
classes of binaries hosting a compact source (figure~\ref{f11}). For IPs,
$P_{\mathrm{spin}} \sim 0.1 P_{\mathrm{orb}}$, while for polars, $P_{\mathrm{spin}} \sim
P_{\mathrm{orb}}$. This is not applicable to $\gamma$ Cas analogs as they are not
Roche-lobe filling systems, unlike IPs. A better analogy is B\textit{e}/NS
binaries. \citet{corbet84} found a relation $P_{\mathrm{spin}} \propto
P_{\mathrm{orb}}^{2}$ in observations (gray symbols in figure~\ref{f11}), which was
later explained by \citet{waters89} in theory; if the binary separation is larger, hence
the $P_{\mathrm{orb}}$ is longer, the density of the B\textit{e} star wind or decretion
disk is smaller close to the NS. As a result, a smaller mass accretion rate is achieved
by the Bondi-Hoyle accretion \citep{bondi44}, and a slower spin is needed to balance
with the in-falling ram pressure.

\citet{apparao02} argued that the same applies to B\textit{e}/WD binaries, though they
used $P_{\mathrm{spin}}$ of $\gamma$ Cas that is considered to originate from the
B\textit{e} star rotation \citep{smith16}. Still, the analogy holds by scaling the
relation from NS to WD as
\begin{equation}
 P_{\mathrm{spin}} \propto P_{\mathrm{orb}}^{2}
 \left(\frac{M_{\mathrm{WD}}}{M_{\mathrm{NS}}}\right)^{-11/7}
 \left(\frac{R_{\mathrm{WD}}}{R_{\mathrm{NS}}}\right)^{18/7}
 \left(\frac{B_{\mathrm{WD}}}{B_{\mathrm{NS}}}\right)^{6/7},\label{e08}
\end{equation}
where $M_{\mathrm{X}}$, $R_{\mathrm{X}}$, and $B_{\mathrm{X}}$ respectively are the
mass, radius, and the magnetic field strength of WD (X=WD) or NS (X=NS). The scaled
relation is shown in red in figure~\ref{f11}. With the observed $P_{\mathrm{orb}}$ of
$\gamma$ Cas, we would expect its $P_{\mathrm{spin}}$ to be several times of 10~ks. This
is not easy to detect with typical $\lesssim$50~ks observations using observatory-type
X-ray telescopes. Two exceptions are a very long (54~hr), nearly continuous observation
with the Rossi X-ray Timing Explorer \citep{robinson00} and the eight-year monitoring by
the all-sky monitor such as the Monitor of All-sky X-ray Image (MAXI; \cite{matsuoka09}). No
significant detection of a coherent X-ray pulse was reported. We need to obtain a data set
of relatively short but high cadence exposures with a sampling rate that matches the
targeted frequency range for $\gamma$ Cas and other analogs to argue for or against the
magnetic cases.

\subsubsection{X-ray outbursts}\label{s4-3-2}
In B\textit{e}/NS binary systems, three major types of X-ray temporal behaviors are
known: (1) persistent low-luminosity ($\sim$10$^{34}$~erg~s$^{-1}$) emission, (2) type I
outbursts associated with an orbital motion with emission elevated by 10$^{2}$--10$^{3}$
times, and (3) type II outbursts with even larger emission but not associated with
orbital periods. The different behavior is well understood as a dynamical interaction
between the B\textit{e} decretion disk and the NS, as is shown in a series
of smoothed particle hydrodynamic simulations. The B\textit{e} disk is effectively
truncated at a tidal resonance radius where the mass accumulates \citep{osaki96}. The
persistent emission is due to the mass leaking from the truncated disk, type I outbursts
occur when the NS passes close to the disk at the periastron, and type II outbursts
occur when the disk is drastically disturbed \citep{okazaki02,okazaki13}.

The theory explains various observations very well, in particular, the presence of type
I outbursts in some systems and the absence thereof in others. In systems with a low
eccentric orbit, which is represented by X Per, the disk is truncated at a 3:1 resonance
radius. The gap between the truncation point and the first Lagrangian point of the
binary is so large that the elevated mass accretion at periastron (type I outbursts)
does not take place, unlike systems with a highly eccentric orbit \citep{okazaki01}. If
the underlying physics is the same, the theory also applies to the B\textit{e}/WD
systems. It is therefore quite important to verify if there is any behavior similar to
type I outburst of B\textit{e}/NS binaries. If none is found, this would suggest that
the $\gamma$ Cas analogs have a wide and low ($\lesssim$ 0.3) eccentric orbit.

\subsubsection{Dwarf or classical novae}\label{s4-3-3}
Yet another interesting clue is dwarf or classical novae. If a $\gamma$ Cas analog hosts
a non-magnetic WD, we would expect changes in the X-ray spectral shape associated with the
changes of the mass accretion rate. In general, the X-ray spectra become softer when the
accretion rate increases during outbursts because the boundary layer emission becomes
optically thick to itself \citep{pringle79}.

Classical novae would occur if a sufficient mass is accumulated on the WD surface to
ignite a thermonuclear runaway. For $\gamma$ Cas and HD\,110432, the lack of classical
nova detection to date is reasonable considering their low $M_{\mathrm{WD}}$ and
$\dot{M}_{\mathrm{X}}$ values. \citet{kato14} presented the minimum
accreted mass to ignite an explosion as a function of $M_{\mathrm{WD}}$ and the mass
accretion rate ($\dot{M}$). For $M_{\mathrm{WD}} \sim 1 M_{\odot}$ and $\dot{M} \sim
10^{-10} M_{\odot}$~yr$^{-1}$, $>$0.3~Myr is required to accumulate the ignition mass,
which is too long to be observed by chance. If there are any $\gamma$ Cas analogs with a
much higher $M_{\mathrm{WD}}$ and $\dot{M}_{\mathrm{X}}$, the chance of detection would
be less desperate.


\section{Summary}\label{s5}
A debate is on-going for the X-ray production mechanism of the $\gamma$ Cas analogs. We
adopt the B\textit{e}/WD scenario as a working hypothesis and applied the physical models
developed for classical accreting WD binaries to the X-ray spectra of two representative
$\gamma$ Cas analog sources using the Suzaku and NuSTAR data in a wide energy range.

We found that both sources are reasonably explained by the models of non-magnetic or
magnetic accreting WD binaries with the physical parameters in a range consistent with
the classical WD binaries. We investigated the additional spectral evidence, but none of
them was conclusive enough to classify the two sources into either non-magnetic or
magnetic accreting WD binaries. Further observations, especially long-term X-ray
temporal behaviors, are useful to understand the nature of these sources.

We should note here that there is no reason that all $\gamma$ Cas analogs belong to the
same sub-class of accreting WDs: some $\gamma$ Cas analogs may host magnetic WDs while
others do non-magnetic WDs. The present study should be expanded to other $\gamma$ Cas
analogs to investigate their diversity with respect to classical WD binaries.

We intentionally did not argue for or against the other scenario --magnetic B\textit{e}
star-- for $\gamma$ Cas analogs. However, we hope that the presented result by our
approach advances the debate by (1) facilitating discussion to focus on more specific
questions such as how the observed $\dot{M}_{\mathrm{X}}$ can be achieved and (2)
motivating observations that are useful to further understanding of the system.

\medskip

The authors express gratitude for the anonymous reviewer for many insightful comments
that improved the draft significantly. We also appreciate Takeshi Shionome, Misaki
Mizumoto, Yasuharu Sugawara, Reiho Shimomukai, and Tae Furusho at JAXA ISAS and Hirokazu
Odaka at the University of Tokyo for help in data reduction, and Atsuo T. Okazaki at
Hokkai Gakuen University for discussion on B\textit{e}/NS binary systems. This work was
supported by the Hyogo Science and Technology Association, the JSPS Grants-in-Aid for
Scientific Research JP17K18019, JP16K05309, and JP15H03642, and Grant-in-Aid for
Scientific Research on Innovative Areas JP24105007. This research made use of data
obtained from Data ARchives and Transmission System (DARTS), which provided by Center
for Science-satellite Operation and Data Archives (C-SODA) at ISAS/JAXA and from the
European Space Agency (ESA) mission \textit{Gaia}
(\url{https://www.cosmos.esa.int/gaia}), processed by the \textit{Gaia} Data Processing
and Analysis Consortium (DPAC,
\url{https://www.cosmos.esa.int/web/gaia/dpac/consortium}). Funding for the DPAC has
been provided by national institutions, in particular, the institutions participating in
the \textit{Gaia} Multilateral Agreement.

\bibliographystyle{aa}
\bibliography{ms}
\end{document}